# Road Slope Prediction and Vehicle Dynamics Control for Autonomous Vehicles

Gautam Shetty, Sabir Hossain, Chuan Hu* and Xianke Lin*

*Abstract*— **Autonomous vehicles can enhance overall performance and implement safety measures in ways that are impossible with conventional automobiles. These functions are executed through vehicle control systems, which have been the subject of considerable research. Autonomous cars have a distinct advantage as they possess various perception sensors that can predict road surface conditions and other phenomena ahead of time. Many modern automotive control systems treat the road slope as a constant and don't account for changes in the road profile in their vehicle models. As a result, vehicle states may be miscalculated, which, in the worst-case scenario, may result in accidents. This is particularly true for high center-of-gravity vehicles like trailers and delivery trucks. With the help of perception sensors in autonomous vehicles, a road slope estimation system can be developed to aid these control systems in making informed decisions regarding the vehicle's state. The current review is divided into three logical steps that can be discussed in the following manner: the first section describes and reviews the individual steps for developing a road slope estimation system. The second one provides a detailed review of previous investigations that implemented different methods that employ this prediction system to improve overall vehicle performance. Finally, a roll control system is presented as an innovative idea that builds on the whole discussion. A rollover prevention system with prediction abilities is presented because (1) it proves to be a critical safety feature, especially for heavy vehicles like buses, trucks, delivery trailers, etc., and (2) not enough research has been conducted on technologies that integrate a roll stability controller with a slope estimation system.**

## I. Introduction

A current trend in the automotive industry is the introduction and implementation of driverless cars [1]. As autonomous vehicles (Avs) aid in the creation of accessible systems for those in need and those who are unable to drive, this has been viewed as a positive development. They help in lowering the price and duration of transportation systems [2]. Furthermore, by preventing fatal collisions, enabling elderly and disabled people to move around more easily, expanding the capacity of the road, conserving fuel, and reducing emissions, Avs have the potential to alter transportation systems fundamentally. Vehicles may become an on-demand service as a result of trends in shared transportation and ownership that are complementary. Transportation preferences, parking requirements, trucking, and other activities may all be impacted. The passenger compartment could change: former drivers might safely use it to work on laptops, eat meals, read books, watch movies, and/or make phone calls to friends [3]. Numerous AV technologies have seen significant advancements in bringing Avs into real-world applications in laboratory tests, closed-track tests, and public road tests. Many stakeholders, including transportation organizations, IT behemoths (like Google, Baidu, etc.), transportation networking firms (like Uber, DiDi, etc.), automakers (like Tesla, General Motors, Volvo, etc.), chip and semiconductor manufacturers (like Intel, Nvidia, Qualcomm, etc.), and so on, have made sizable investments in and promoted these advancements [4]. Fig. 1 lists the specifications of Avs by different companies for the year 2020.

The Defense Advanced Research Project Agency (DARPA) Grand Challenge, which was held in 2007 and evaluated autonomous navigation technologies for urban environments, marked a significant advancement in autonomous driving technology. The majority of the winners focused on environment perception, precise localization, and navigation in order to perform a variety of urban driving manoeuvres, such as lane changes, U-turns, parking, and merging into moving traffic [5]. Autonomous driving is achieved by the cyclic action of three steps, namely environment sensing, perception, decision and action. The perception domain consists of environment sensing, mapping and localization, sensor fusion and V2X communications. Consequently, the planning/decision consists of various plans required to calculate the optimal motion path. Finally, vehicle control includes mechanisms to achieve these virtual decisions in real-life [6]. The most up-to-date driving systems, from the earlier advanced driver assistance system (ADAS), to the adaptive cruise control (ACC) system, and to the latest combined active front steer (AFS) and direct yaw control (DYC) system, have been successfully deployed in the

This work was supported by the Dr Xianke Lin's startup funding (Corresponding author: X. Lin, C. Hu).

Gautam Shetty is Department of Automotive and Mechatronics Engineering, Ontario Tech University, Oshawa, ON L1G 0C5, Canada (e-mail: gautam_s@uoit.ca).

Sabir Hossain is with Department of Automotive and Mechatronics Engineering, Ontario Tech University, Oshawa, ON L1G 0C5, Canada (e-mail: sabir.hossain@ontariotechu.net).

Chuan Hu is with School of Mechanical Engineering, Shanghai Jiao Tong University, 800 Dongchuan Road, Shanghai 200240, China. (chuan.hu.2013@gmail.com*)

Xianke Lin is with Department of Automotive and Mechatronics Engineering, Ontario Tech University, Oshawa, ON L1G 0C5, Canada (xiankelin@ieee.org*).

Color versions of one or more of the figures in this article are available online at http://ieeexplore.ieee.org

intelligent vehicles. Those associated technologies can comprehensively enhance the enormous potential of vehicle dynamics capacity in safety and handling stability [7]. The accuracy of path tracking has drawn more attention in the current research on autonomous vehicle control. However, vehicle stability was frequently disregarded in path tracking

- o Development of Vehicle Dynamics models
- o Control systems based on longitudinal slope
- o Control systems based on lateral slope
- An innovative idea - Road slope integrated anti-roll control system

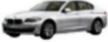

| Many carmakers are developing prototype vehicles that are capable of driving autonomously in certain situations | | |
|---|---|---|
| **BMW** | **Mercedes-Benz** | **Google** |
| **Vehicle** | | |
| 5 Series (modified) | S 500 Intelligent Drive Research Vehicle | Prius and Lexus (modified) |
| **Key Technologies** | | |
| • Video camera tracks lane markings and reads road signs<br>• Radar sensors detect objects ahead<br>• Side laser scanners<br>• Ultrasonic sensors<br>• Differential GPS<br>• Very Accurate map | • Stereo Camera Sees objects ahead in 3-D<br>• Additional cameras read road signs and detect traffic lights<br>• Short- and long- range radar<br>• Infrared camera<br>• Ultrasonic sensors | • LIDAR on the roof detects objects around the car in 3-D<br>• Camera helps detect objects<br>• Front and side eradar<br>• Inertial measuring unit tracks position<br>• Wheel encoder tracks movement<br>• Very accurate map |

Fig. 1. Specification of Avs by the year 2020 adapted from [8]

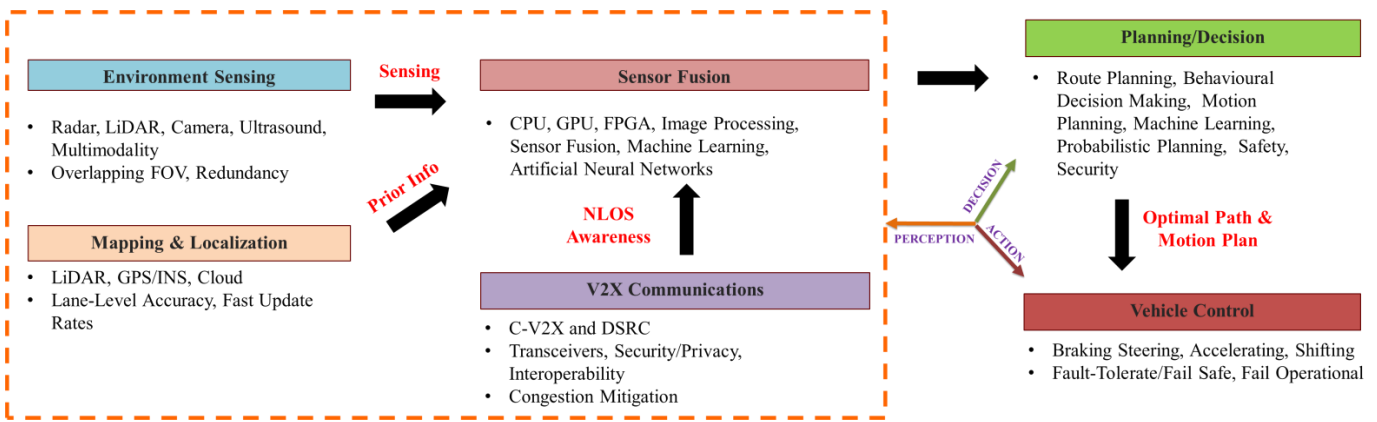

Fig. 2. Classification, Connotation, and Interrelation of Autonomous Driving Technologies adapted from [6]

control research. The target path is tracked, particularly in some commercial vehicles, but in some extreme circumstances, yaw stability or roll stability are threatened. Even if the accuracy satisfies the requirements, path tracking without a security guarantee is useless [9]. Fig. 2 outlines the basic technologies that form a self-driving system and their interrelations.

The topics tackled within this review are: -

- Development of a road slope estimation system through sensors
- Vehicle control systems that account for road slope

This paper will describe and review various steps in developing a road slope estimation system. This is an important section because the estimation process is an integral part of the technique as it alarms the control system to prepare the vehicle for sudden road slopes or grade changes. Without this system, the car may be unable to adjust its speed or head in time to prevent rollover, potentially leading to catastrophic failure.

After that, various control systems are discussed that make use of the road slope data gathered from the previous program. Before going into the details of the control system, it is necessary first to describe the vehicle dynamics behind it. This part consists of a tyre and longitudinal vehicle models that account for slope. This discussion and review are essential because it was found, through this review, that road slope integrated control systems can lead to fuel-efficient driving strategy, intelligent vehicle control and pose estimation systems.

Table I
COMPARISON OF VARIOUS PERCEPTION SENSORS

| Criterion | Lidar | Stereo Camera | Monocular Camera | Depth Camera |
|---|---|---|---|---|
| Usage | Commonly used to make high-resolution maps. Creates an accurate long- and short-range map of a vehicle's surroundings | Used for depth estimation, object detection, and semantic segmentation. Also, supplement Lidar for Better Automotive Advanced Driver Assistance Systems. | Usually placed on every side — front, rear, left and right – of a self-driving car to create a 360 degree view of the surrounding environment. | Quite functional in the field of autonomous devices and robotics like drones for collision avoidance |
| Working | Time of Flight method - LiDAR emits a laser pulse, and the sensor returns and records the reflected light energy. | Uses binocular vision to recover a 3D structure of a scene | Reflection of light to create an image | Time of flight method |
| Depth Accuracy | 3% relative error at a distance of 300 meters | Generally, 1 - 2mm at up to approximately 2 metres | Not Applicable | The average accuracy of 0.5mm and at a closer range within 0.2 meter |
| Weather Cond. | Often unreliable at night-time or in inclement weather | Work great in almost every lighting condition including outdoor environments | Work great in almost every lighting condition including outdoor environments | Often unreliable at night-time or in inclement weather |
| Cost (USD) | ~$8,000 | ~$600 | ~$50 | ~$250 |
| Criterion | Lidar | Stereo Camera | Monocular Camera | Depth Camera |

Ultimately, an innovative idea based on the entire review is presented. This idea combines the concepts discussed before it, namely, the road slope estimation system and the vehicle dynamics control systems. It can be termed as a rollover prevention system which uses slope information. An anti-roll control system is chosen for this analysis because it proves to be a critical safety feature, especially for heavy vehicles like buses, trucks, delivery trailers, etc., and its functionality can be elevated by considering the road slope. Furthermore, not enough research has been conducted on technologies that integrate a roll stability controller with a slope estimation system.

The previous list forms a logical path followed within this paper to explain and review different aspects of the problem to be tackled. This paper is necessary because it reviews a topic that can be a further leap in the field of autonomous vehicle navigation.

The structure of this paper is organized as follows: Section II reviews the road grade and cross-slope estimation methods using different kinds of perception sensors. Section III presents the development and review of a vehicle model for the controller development. Section IV reviews various vehicle control methodologies that account for road slope changes. The future prospects are given in Section V. And finally the conclusion is given in Section VI.

## II. ROAD SLOPE ESTIMATION THROUGH SENSORS

The above slope estimation process (Fig. 3) is adapted from [10]; some steps have been omitted so that they can fit a broader category of strategies employed by other research studies. [11] and [12] follow a similar approach with some modifications. Both of these papers do not perform the drivable space estimation method and instead directly calculate the road slope for a fixed path in front of the vehicle. On the other hand,

[13] proposes a way that does the process in reverse order where the calculated longitudinal profile through disparity maps is used to find the drivable space. Several other studies for vision-based slope estimation do not follow the above pipeline to generate elevation profiles. For example, describes four methods using geometry and deep learning techniques to find the required slope.

### A. Road/Sidewalk Visual Capture

Road grade (longitudinal slope) and cross-slope (lateral slope) is the first step in a roll stability control system built for autonomous vehicles. Autonomous vehicles use perception sensing to gather data from their sensors and translate that data into an understanding of their surroundings [14]. The perception system may consist of a Lidar, monocular camera, stereo camera, and (or) a depth camera. These devices are generally attached to a single robot where each sensor performs a particular function. Lidar is used to generate high-definition maps during testing phases. Monocular cameras generally perform object detection and tracking. A camera with two or more lenses and a separate image sensor or film frame for each lens is referred to as a stereo camera. As a result, the camera can replicate human binocular vision and produce three-dimensional images. These cameras are used for object detection, depth estimation, and semantic segmentation. Depth cameras are also able to estimate the depth of the surroundings. Table I compares the different perception sensors i.e. Lidar, Stereo Camera, Monocular Camera and Depth Camera on various criteria.

| Method | KITTI 2012 | | KITTI 2015 | | Time |
|---|---|---|---|---|---|
| | Out-Noc | Out-All | D1-bg | D1-all | |
| PSMNet [16] | 1.49 | 1.89 | 1.86 | 2.32 | 0.41 |
| GA-Net [17] | 1.36 | 1.80 | 1.55 | 1.93 | 1.5 |
| DispNetC [18] | 4.11 | 4.65 | 4.32 | 4.34 | 0.06 |
| AANet [15] | 1.91 | 2.42 | 1.99 | 2.55 | 0.062 |
| GC-Net [19] | 1.77 | 2.30 | 2.21 | 2.87 | 0.9 |

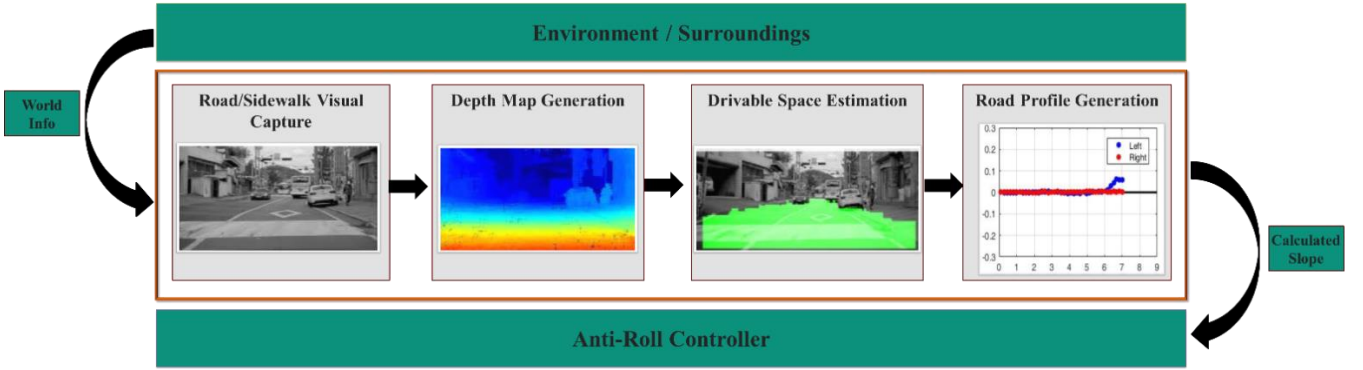

Fig. 3. Road Slope estimation process adapted from [10]

Russell A. White et al. have used LiDAR-derived digital elevation models to perform road mapping on dense forest roads. A 1 m DEM Digital Elevation Model was used to precisely extract the location, gradient, and length of a forest haul road in the Santa Cruz Mountains, California [20]. In the form of 3D point clouds, highly dense, irregularly distributed georeferenced data can be rapidly collected using mobile LiDAR. This technology has become more and more popular for identifying roads and other road-scene objects [21]. For monocular cameras, Apoorva Joglekar et al. present a novel approach which uses road geometry and point of contact on the road to estimate the distance of any vehicle [22]. Turgay Senlet et al. create a framework of vehicle localization by reconstructing roads on the Rutgers University campus through stereo cameras. Stereo image pairs for various routes are obtained, depth images and VO are calculated from them, and then they are converted to frame point clouds using projective geometry [23]. Teng Cao et al. also use a similar concept of calculating the depth image from the stereo data, but they employ a novel slope analysis method called the V-intercept, which can detect slopes and obstacles directly from the disparity map. Thus, they create a novel navigation framework for autonomous vehicles equipped with stereo cameras, featuring finishing all the perception and path planning tasks [24]. Finally, a depth camera in the form of a Kinect sensor was used by F. Marinello et al. to analyze forest road surface roughness of gravel roads [25]. Even though the results are less detailed and accurate than the Lidar method [14], a Kinect sensor is much cheaper (less than 100 €) than a brand new lidar sensor and thus can be used for budgeted operations.

### B. Depth Map Generation

After obtaining the images from the sensors, the next step is to perform road surface profile estimation to get the required slopes. The best way to acquire spatial information from an image or a set of images is to obtain its depth/disparity map. In computer graphics and computer vision, a depth map is an image or image channel that contains information about the distance between the surfaces of scene objects from a particular vantage point. This information can be further used to reconstruct the captured 2D scene into a 3D model in the global coordinate reference frame.

The most popular method for estimating depth in robotics and computer vision is stereo vision [26]. Even though Lidar systems are quite powerful, they are extremely expensive (the average price of vehicle-mounted semi-solid LiDAR is about $8,000), which renders them incapable of being used on production automotive perception systems. Also, modern stereo camera systems can catch up to lidars, for example, stereovision can deliver approximately 2,000 vertical samples per second

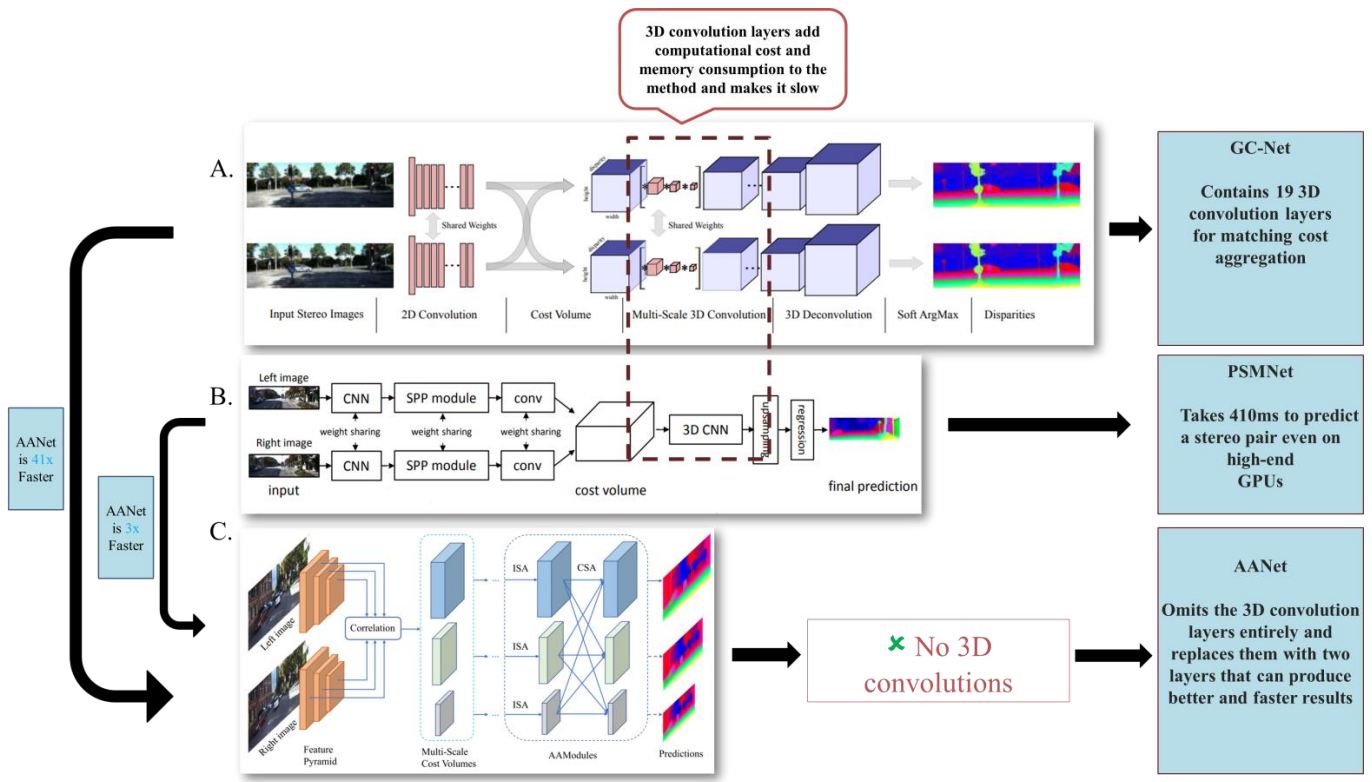

Fig. 4. Methodology comparison of various depth map generation methods. Here processes (A) and (B) denote the methodology employed by GC-Net [19] and PSMNet [16] respectively. Process (C), which is AANet [15] presents a better way by completely avoiding 3D convolution layers, thus making it significantly faster than the other two methods

using today's generation of cameras. In comparison, LiDARs can provide only 128 vertical samples per second, clearly showing that stereovision can deliver superior resolutions. Stereo resolution is also strong in low-light scenarios [27]. In 2019 researchers at Cornell University [28] presented a report where a method for detecting objects using inexpensive cameras achieved almost the same accuracy as the LiDAR. This shows that, in theory, it is possible that stereo cameras could replace LiDAR in the future. Finally, coming on to depth cameras, even though they can provide depth maps directly, depth calculation is reliable for indoor applications only. Also, since stereo cameras use visual features to measure depth, they will work well in most lighting conditions, including outdoors.

Going over depth map calculation methods for stereovision, one can use compute vision algorithms like Stereo Global Block Matching (SGBM). SGBM is a modified version of the Semi-Global Matching (SBM) algorithm that was proposed by Hirschmuller [29], [30] This method can be easily implemented in code through the OpenCV Library. Next, the WLS Filter which is also available in OpenCV library can be used to optimize the output of stereo matching [31]. There are also deep learning-based methods to calculate required depth maps. Tyler S. Jordan et al. explore the benefit of using Convolutional Neural Networks (CNN) for rendering disparity maps from stereo imagery. Particularly in areas with little to no texture, this technique can significantly outperform the naive subtractive plane-sweep procedure used in conventional algorithms [32]. The first end-to-end trainable disparity estimation framework,

DispNetC [18], uses a correlation layer to assess how similar the left and right image features of a stereo pair are to one another. Then, by directly concatenating the left and right features, GC-Net [19] adopts a different strategy, necessitating the use of 3D convolutions to aggregate the resulting 4D cost volume. PSMNet [16], which offers more 3D convolutions for cost aggregation and thereby achieves better accuracy, further enhances GC-Net. Although 3D convolutions can deliver cutting-edge performance, they are quite expensive to use in practise due to their high computational cost and memory consumption. The limitations of 3D convolutions are also recognised by GA-Net [17], who attempts to replace them with two guided aggregation layers, but their final model still utilises fifteen 3D convolutions. To mitigate these issues, AANet [15] designed two modules for cost aggregation, thus eliminating the 3D convolution layers. These two modules are lightweight and simple, leading to an efficient architecture for cost aggregation resulting in a highly accurate model. Fig. 4 compares the methodologies adopted by the 3 different (GC-Net [19], PSMNet [16], AANet [15]) depth map generations technologies. Table II compares the benchmark results for 5 different depth map generation technologies (DispNetC [18], GC-Net [19], PSMNet [16], GA-Net [17], AANet [15]) for both the KITTI 2012 and 2015 datasets [33].

While these models tend to perform better than plain mathematical algorithms, there is a recurring issue of generalization. Generalization is the capacity of a model to fit correctly to novel, previously unobserved data taken from the same distribution as the model's training data [34]. But there have been methods to mitigate this issue. Jiahao Pang et al. from SenseTime research proposed a method named zoom and learn, or ZOLE for short, to resolve this dilemma. They present a self-adaptation approach to generalize deep stereo matching methods to novel domains. The model utilizes synthetic training data and stereo pairs of the target domain, where only the synthetic data have known disparity maps [35].

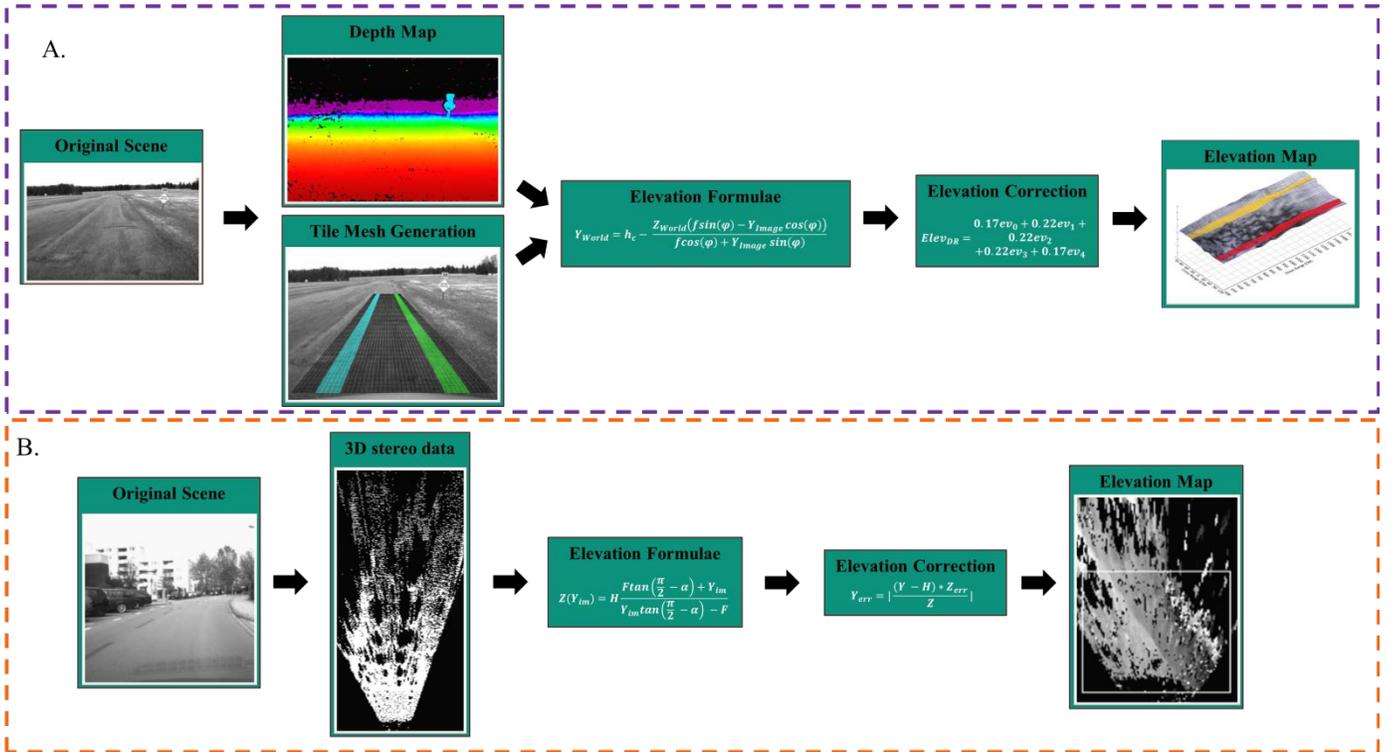

Fig. 6. Road Profile Generation Process. Here process (A) denotes the methodology employed by [11] where a depth map and tile mesh are used for elevation profile generation and (B) denotes a similar method used by [39] but it uses a different formula for elevation and elevation correction and also it works on 3D stereo data rather than depth map. The white rectangle represents the region of interest in the last image in 3D.

For computing depth maps of monocular images, Bo Li et al. demonstrate competitive results by using regression on deep convolutional neural network (DCNN) features, in addition to a postprocessing refining step using conditional random fields (CRF) [36]. Even though there are several deep learning technologies to calculate depth maps from monocular images, it is still a challenging and essentially underdetermined problem. Nikolai Smolyanskiy et al. from NVIDIA point out that despite the progress on monocular depth estimation in recent years, the gap between monocular and stereo depth accuracy is still great. This is a important result as many self-driving cars that are currently being developed have a widespread reliance on monocular cameras [37].

*C. Drivable Space Estimation*

Now the regions must be identified where the road surface profile will be calculated. This is termed free space / drivable space estimation. Fig. 5 explains the purpose of free space estimation algorithm. Nicolas Soquet et al. present an original approach to free space estimation by stereovision. First, the road profile is obtained by creating a v-disparity map. This road profile is used along with the u-disparity image to classify pixels of the disparity map as free space. Then finally, a colour segmentation is performed on the free space [38]. Deep learning-based techniques can also be used to achieve this. Two modules make up Mahmoud Hamandi et al.'s fully self-supervised system. The first module is based on a Fully Convolutional Network (FCN). After the system is started, it is used for ground segmentation. The second module trains the FCN at startup and whenever its performance deteriorates over the course of a run. It depends on depth data and interactive graph cuts. This technique can extract reliable free ground space in a stereo image pair without any offline training [40]. Free space estimation methods have also been established for monocular cameras. The entire problem of finding a drivable free space was reduced by Jian Yao et al. to an inference problem on a 1D graph, where each node represents a column in the image and its label designates a position that separates the free space from the obstacles [41]. Koichiro Yamaguchi et al. from Toyota Central R&D Labs., INC. propose a method for estimating the road region in images captured by a vehicle-mounted monocular camera. The method first estimates the camera motion and road plane parameters; then, it calculates a homography matrix for the road plane between two successive images. In the end, the road region is detected by differentiating the warped previous image and the current image [42].

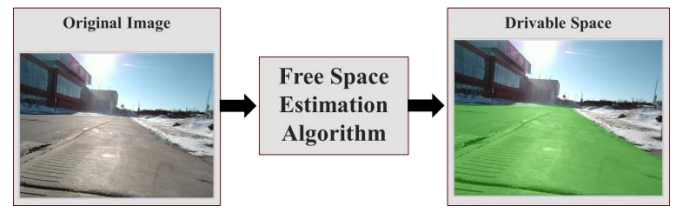

Fig. 5. The above figures explains the function of a free space estimation algorithm. The aim of a drivable space program is to pinpoint the area from the original which is drivable. In the right figure it is correctly shown that green is the drivable space as it consists of the road and sidewalk and the buildings and grass area is omitted from the free green space. The final road profile or slope will be calculated for this green patch.

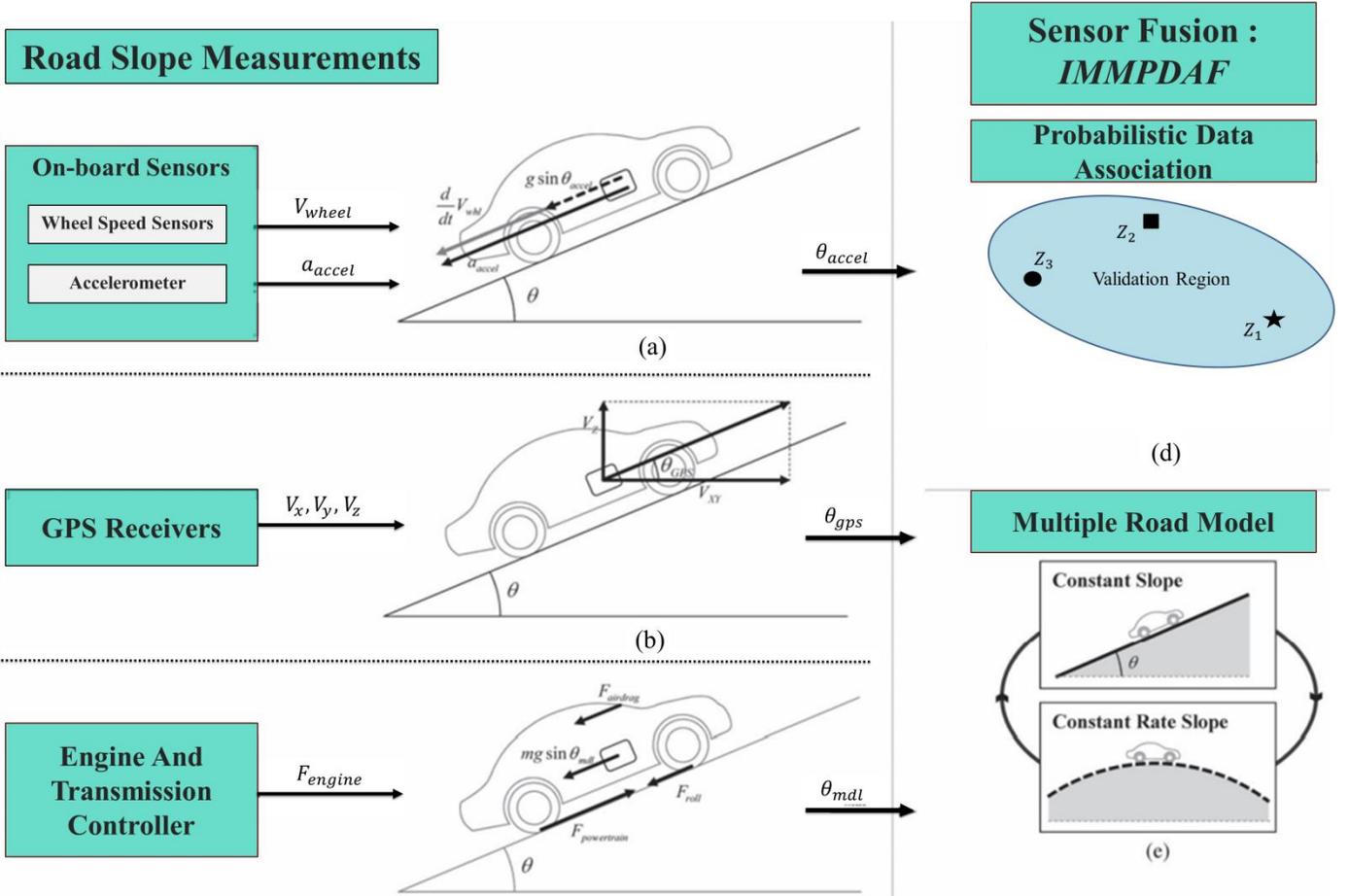

Fig. 7. The overall architecture of the real-time road-slope estimation system is adapted from [43].

### D. Road Profile Generation

The final step is to create the digital elevation map. The process of creating a digital elevation map refers to converting the depth map into a height model where each unit in the free space denotes its height over the ground reference plane. Fig. 6 describes two different but comparable technologies to generate elevation maps. Truman Shen et al. from Takata holdings inc. describe a method where they divide the free space part of the depth map into quantifiable tiles. By grouping measurements within each tile, elevation within each tile (here

it is denoted by $Y_{World}$) can be computed as following equation (1): -

$$Y_{World} = h_c - \frac{Z_{World}\big(f\sin(\varphi) - Y_{Image}\cos(\varphi)\big)}{f\cos(\varphi) + Y_{Image}\sin(\varphi)} \quad (1)$$

Here the parameters are defined as
- $Z_{World}$ is the median downrange per tile
- $Y_{Image}$ is the average image row per tile
- $f$ is the camera's calibrated focal length
- $k_g$ is the stereo camera's mounting height
- $\varphi$ is the stereo camera's mounting pitch angle

Here the elevation is measured relative to a virtual reference surface determined by the stereo camera's extrinsic parameters.

For more accuracy, tile elevation is calculated as a weighted sum of laterally adjacent tiles. The dynamic pitch angle is determined at runtime and the formulae are modified accordingly to account for the stereo camera's various vibrations [11]. A similar method is employed by Florin Oniga et al., but they were able to extend this basic model to calculate the elevation and explained how to use this basic model to compute depth resolution even for nonplanar road surfaces [39]. Another benefit of depth maps is that they can be directly converted to point clouds. These clouds contain 3d coordinates of each and every point within the scene along with the RGB information and they can be used to reconstruct the free space or the road surface. OpenCV's utility for doing this calculation is a function called reprojectImageTo3D, which takes in the disparity map and a matrix representing the camera's intrinsics, including its field of view, focal length, and output resolution. The function returns an output image with each pixel being the 3D coordinates of that corresponding pixel. This method can be used for both stereo image pairs and monocular images, making it a viable strategy.

### E. Other Methods to Calculate Slope

There are several other methods to perform road slope estimation without going through the pipeline discussed till now, but some of these techniques only apply to certain terrains or roads. To go into a little more detail, the methodology that was discussed till now works for any scene, whether it is a structured, forest, mountainous road or a sidewalk because this method does not depend on any particular visual features to calculate the slope. On the other hand, certain techniques are discussed below, which can calculate the longitudinal or lateral slope in a much easier fashion, albeit they only work for specific types of scenes. These techniques are pretty different from each other and thus have to be categorized in a separate section.

Eser Ustunel et al. have presented three vision-based methods to calculate slope using road lines or local features from images. The first is a geometry-based (GB) method; it obtains the slope estimates using 2D road lines derived from 3D road lines and pin-hole camera models. The next is a local features-based (LFB) method, which uses the SIFT [44] local features between two consecutive image pairs. The scale-invariant feature transform (SIFT) is a computer vision algorithm to detect, describe, and match local features in images. And the third solution is a covariance-based (CB) method, which considers supervised learning where 2D road lines are used as features to train the algorithm. GB and CB methods rely on road lines and provide in advance estimations that enable proactive driving. In contrast, LFB uses the entire image and performs estimation when the road slope changes [45]. Another method for structured roads is presented by Yunbing Yan et al. where they describe a technique similar to the CB one [39]; within this, the lane-line feature information in front of the vehicle is obtained according to machine vision, and the lane-line function is fitted according to an SCNN (Spatial CNN) algorithm, then the lateral slope is calculated by using an estimation formula which is the relationship between the lateral road slope and the tangent slope of the lane line that can be found out according to the image-perspective principle; then, the coordinates of the pre-scan point are obtained by the lane line, and the tangent slope of the lane line is used to obtain a more accurate estimation of the road lateral slope [46].

### F. Real-time Slope Estimation Methods

Apart from predicting the slope ahead of time using camera sensors, specific arrangements must be made within the vehicle to calculate the slope in real-time for accurate measurements. Present as well as future data can be used in tandem to help the controller make a safe, conservative estimate to prevent rollover. There are numerous methods to accomplish this. GPS-based estimation methods are the most direct way to do so. In this paper, two methods for calculating road grade using a ground vehicle's GPS system are illustrated. The first method directly measures the vehicle's attitude in the pitch plane using two antennae; the second method estimates road grade using vertical to horizontal velocity at a single antenna [47]. In order to increase the effectiveness and performance of these real-time methods, this paper suggests a combined road-slope estimation algorithm. The algorithm blends three different types of road-slope measurements from a longitudinal vehicle model, an automotive onboard sensor, and a GPS receiver. A probabilistic data association filter (PDAF) is used to accomplish the measurement integration. It assigns a statistical probability to each measurement in order to combine them into a single measurement update and removes inaccurate measurements using the PDAF's false-alarm function. Fig. 7 displays these technologies schematically [43].

## III. Vehicle Model Development

The slope calculated from the previous steps is sent to a controller that uses this information to adjust the vehicle state for improved performance and safety. The following section will outline the development of a vehicle model for a vehicle control system that utilizes slope data. The vehicle model for a longitudinal vehicle state controller consists of two distinct parts: - a tire model and a longitudinal motion model. These two physics-based systems work together to simulate a moving vehicle. Vehicle dynamics models have evolved from the conventional lumped parameter model to the finite element model (FEM), the dynamical substructure model, and the multi-body system dynamics model. They have also gone from the linear model to the non-linear model with the non-linear stiffness and damping [48]. But to keep the calculations reasonably simple, we will be exploring lumped parameter models for our controller.

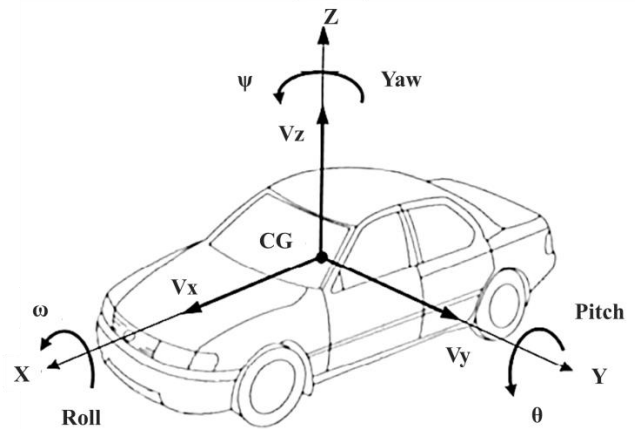

Fig. 8. Vehicle Axis System ISO 8855-2011 adapted from [51]

Fig. 8 depicts the coordinate system that guides a vehicle's movement. The midpoint of the axes ($x$, $y$, and $z$) of motions in the direction of the x-axis is the centre of gravity, also known as the point of gravity. Yaw motion and lateral motion both move in the direction of the y-axis. Pitch motion and vertical motion (oscillating up-down) are z-axis-directed movements

[49] [50].

## A. Tyre Modelling

Tyre modelling is of great importance; after all, it provides the only possibility to analyze and quantify the transfer of forces between the vehicle and the road. It describes the connection between the wheel and the chassis. The process of creating a coherent model that can capture the physics of tyres so that they can be mathematically analysed and evaluated is known as "tyre modelling" [52]. In current vehicle simulator models, the tire model is the weakest and most challenging part to simulate.

Tire forces are separated into a longitudinal force component (braking and driving) and a lateral force component (steering/cornering). A tire's longitudinal slip and slip angle in relation to the road determine the longitudinal and lateral forces that are produced by the tyre [53].

We will go over the simple linear model, dugoff tyre model and the magic formula model briefly as they are the easiest to use and understand models for tyre dynamics.

The simple linear tyre model is quite a simple model but tends to be used often due to its simplicity. For low values of longitudinal slip and lateral slip angle, the following relations (equation (2), (3), (4)) can be considered approximately true

$$F_x = C_x S \qquad (2)$$

$$F_y = C_y \alpha \qquad (3)$$

$$M_z = C_M \alpha \qquad (4)$$

where $S$ is the longitudinal slip, $\alpha$ is the slip angle, $C_x$ is the longitudinal (or braking) stiffness, $C_y$ is the lateral (or cornering) stiffness, $C_M$ is known as the aligning stiffness, $F_x$ is the longitudinal force, $F_y$ is the lateral force and $M_z$ is the aligning moment.

The empirical lateral Dugoff tyre model is a simple way of representation of the force and moments generated by the tyre. The primary expression of the formula is given as the equation (5) [55]

$$F(x) = D\sin(C\tan^{-1}(Bx - E(Bx - \tan^{-1}(Bx))))  \qquad (5)$$

where $x$ stands for either $S$ or $\alpha$, $B = C\alpha/(CD)$ is the stiffness factor, with $C_\alpha = c_1 \sin(2\tan^{-1}(F_z/c_2))$, $c_1$ being the maximum cornering stiffness and $c_2$ the load at maximum cornering stiffness, $D$ is the peak factor (often defined as $\mu F_z$), $C$, $E$ are shape factors, which can also have various definitions. There are multiple other versions of the formula. The added complexity of the Magic formula allows it to capture the effects of other factors on the force generation. Fig. 9 plots a general pacejka tire graph where the different constants and their significance has been displayed. Table III compares the 3 fundamental tyre models discussed previously on different

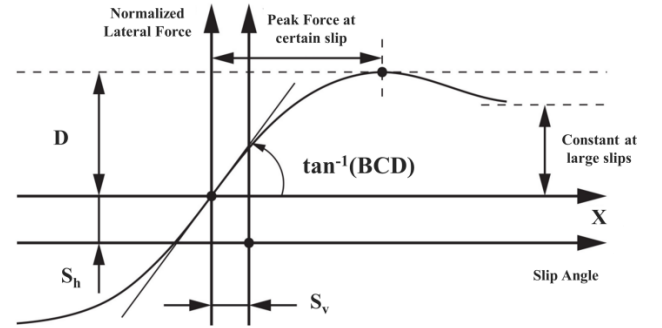

Fig. 9. Pacejka's Magic Formula tire model. The graph shows slip angle on the y-axis vs. lateral force on the Y-axis. Parameters $B$, $C$, $D$, $E$ have been explained previously. atan ($BCD$) is the angle made by the force slip curve at low slip angles. $S_h$ is the horizontal shift and $S_v$ is the vertical shift adapted from [57].

criteria. The above discussion of the three tyre models has been adapted from [56].

Table III
BASIC TYRE MODEL COMPARISON

| Comparison | Tyre models | | |
| --- | --- | --- | --- |
| | Simple linear model | Dugoff model | Magic Formula model |
| **Advantages** | Straightforward to implement | Easy to implement. Can simulate for moderate slip values | Can capture tyre behaviour at a very large slip range. Takes into account the effect of multiple parameters on tyre dynamics |
| **Disadvantages** | Works approximately only for small slip ratios and angles | Deviates drastically from correct behaviour at higher slips | Difficult to implement. Requires a large amount of data and parameters to function accurately |

capturing tyre physics at moderate slip rates. The friction coefficient is set to $\mu = 1$, which is a reasonable approximation of the friction between the vehicle's and the road. For a detailed insight of the Dugoff model refer to this paper [54].

The magic formula model gives a more accurate

## A. Longitudinal Vehicle Model

The above Fig. 10 contains two individual figures, fig. 10 (a) describes the resistance forces that act on a car going uphill and fig. 10 (b) describes the free body diagram of a rotating wheel.

The longitudinal vehicle dynamic model is simply based on the dynamics of the vehicle that generate forward motion. According to Newton's second law, the longitudinal dynamics of a car can be modelled through a point mass model as follows (equation (6), (7), (8), (9)) [58].

$$F_t = M\frac{dv}{dt} + F_a + F_g + F_r \qquad (6)$$

$$F_g = Mgsin(\theta) \qquad (7)$$

$$F_a = \frac{1}{2}\rho C_d A v^2 \qquad (8)$$

$$F_r = MgC_r cos(\theta) \qquad (9)$$

Here the parameters are defined as,
- $M$ is the total mass of the vehicle
- $F_t$ is the traction force
- $F_a$ is the aerodynamic drag
- $F_g$ is the downgrade force
- $F_r$ is the rolling resistance force
- $\rho$ is the air density
- $C_d$ is the drag coefficient
- $C_r$ is the rolling resistance coefficient
- $A$ is the vehicle cross sectional area
- $dv/dt$ is the vehicle acceleration/deceleration.

In EVs (electric vehicles), an electric motor supplies power to the wheels through the transmission that is connected to the electric drive through a gearbox. This system can be modelled under the no losses condition by [59] through the equations (10) and (11).

$$T_g = k_g T_e \qquad (10)$$

$$\omega_g = k_g \omega_e \qquad (11)$$

Here the parameters are defined as,
- $T_e$ is the electric drive torque
- $T_g$ is the gear torque
- $\omega_e$ is the angular speed of the engine
- $\omega_g$ is the angular speed of the geare
- $k_g$ is the ratio of the gearbox

The electric machine, whose model can be static or dynamic, produces the electric torque [61]. Whether a machine is an AC (synchronous or asynchronous) motor or a DC (synchronous) motor, the dynamic model will vary. It explains the connection between the input current and voltage and the electric torque produced. The static model is based on the electric drive efficiency map and is data-driven [62], given by the equations (12) and (13).

$$T_e = T_{e-ref} \qquad (12)$$

$$I_b = \frac{T_e \omega_g}{u_b \eta^k} \qquad (13)$$

Here the parameters are defined as,
- $T_{e-ref}$ is the reference electric torque through efficiency map
- $u_b$ is the battery voltage
- $I_b$ is the battery current
- $\eta$ is the motor efficiency

The vehicle moves forward thanks to the wheel's conversion of the gear torque into traction force. As a result, the vehicle's velocity is determined by the wheel's angular speed in the manner shown below through equations (14) and (15).

$$F_{t/b} = \frac{1}{R_w}T_{g/b} \qquad (14)$$

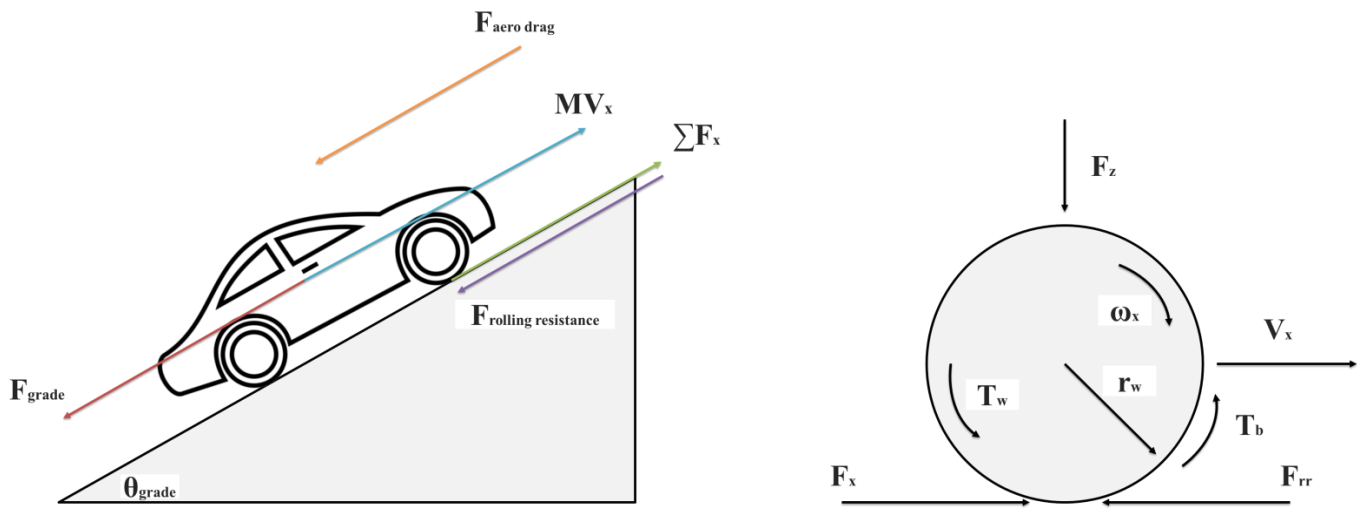

Fig. 10. (a) Longitudinal vehicle dynamics model and (b) simplified wheel dynamics model adapted from [60].

$$\omega_w = \frac{1}{R_w}V \qquad (15)$$

Here the parameters are defined as,

- $F_{t/b}$ is the tractive/braking force
- $T_{g/b}$ is the tractive/braking torque
- $R_w$ is the wheel radius
- $V$ is the vehicle longitudinal velocity

Tire models have been discussed previously and any one of three models namely, simple linear, Dugoff and magic formula model can be used to simulate the forces and moments generated due to tire deformations.

The car's brakes can be either a disc or a drum system. The disc braking system is assumed to be applied to both the front and back wheels so a factor of 4 is included. The braking torque delivered is given as the equation (16).

$$T_b = \frac{\mu P \pi B_a^2 R_m N_{pads}}{4} \qquad (16)$$

Here the parameters are defined as,

- $\mu$ is the brake pad friction coefficient
- $P$ is the applied brake pressure
- $B_a$ is the brake actuator diameter
- $R_m$ is the mean brake pad diameter
- $N_{pads}$ is the number of brake pads

The above discussion on the longitudinal vehicle model is adapted from [63].

## IV. USE OF ROAD SLOPE FOR DYNAMICS CONTROL

We will now review the studies that use this estimated slope to improve vehicle performance. Some of these studies presume that the slope is given. In this review, we assume the road slope information is calculated from the on-board vision system through the process described previously. These studies can be mainly divided into two sections, those that employ longitudinal slope information and those that use lateral slope ones for dynamics control of the vehicle. Those that employ both will be discussed in the end.

### A. Longitudinal Slope Control

Longitudinal slope information can be used for various purposes. This can include developing a fuel-efficient driving strategy, intelligent vehicle control and pose estimation systems, etc. The papers discussed below assume that the slope data is available through vision-based or real-time estimation methods. The slope information obtained through a vision system can be used to adjust the trajectory or vehicle state in advance.

Eco-driving is a way of maneuvering a vehicle by a human driver intended to minimize fuel consumption while coping with varying and uncertain road traffic by trading off the most efficient driving point of the vehicle whenever necessary. Fuel consumed in a car is greatly influenced by road gradients besides its velocity and acceleration characteristics. The following studies will utilize this fact to reduce fuel consumption by dynamically changing the vehicle state through various controllers that account for road slope. Many of these controllers are based on the MPC technique (Model Predictive Control). For example, This paper presents a non-linear model predictive control method with a quick optimization algorithm to determine the vehicle control inputs based on the road gradient conditions obtained from digital road maps. Also, to ensure fuel-efficient driving, the fuel consumption model of a typical vehicle is created using engine efficiency characteristics [64]. [65] employs the model predictive control technique to calculate optimal vehicle control inputs using both traffic signal and road slope information. Information on the traffic signal switching enables the reduction of unnecessary vehicle accelerations and decelerations and the use of the high-efficiency points of the engine and the regenerative braking energy. The outcomes of computer simulations were used to evaluate how well the suggested method performed. The suggested method optimises both the fuel efficiency and the driving style. The development of a non-linear model-based predictive controller for Eco-cruise in autonomous ground vehicles is the goal of this work. An NMPC (nonlinear model-based predictive controller) is used by [66] to reduce fuel consumption and improve safety and comfort levels during a trip. Using latitude-longitude information and a GPS module, the predictive controller calculates a sequence of control inputs to smooth the vehicle's acceleration and brake along the route in critical parts, such as uphills, downhills and curves following the speed limits of each road. The fact that the proposed method only needs publicly accessible data to obtain the controller parameters is a significant advantage. On the other hand, several other studies utilize methods other than mpc or nmpc to create fuel-efficient systems. This study introduces an Adaptive Neuro-Fuzzy Inference System (ANFIS) based on ACC (Adaptive Cruise Control) systems, which lowers the vehicle's energy use and boosts efficiency. Using road data, it determines the vehicle's energy consumption under various dynamic loads, including wind drag, slope, kinetic energy, and rolling friction. Based on the predetermined speed and the predicted future slope information, the cruise control system adaptively regulates the vehicle's speed. The evaluation's findings show that the look-ahead methodology, which was based on the driving cycle, effectively controlled the vehicle's speed, resulting in a 3 percent reduction in average fuel consumption [67]. Finally, the new dynamic programming–based eco-driving algorithm by [68] generates the optimal velocity trajectory using the information on the roadway gradients, longitudinal vehicle dynamics, and vehicle transient fuel consumption characteristics. A newer and more accurate instantaneous fuel consumption model is developed and used in the objective function to ensure fuel economy driving. The simulation results show that approximately 6.89%–24.78% of fuel can be saved using the dynamic programming–based eco-driving algorithm compared with using the cruise control algorithms like the one discussed previously.

Here we will discuss the development of various control systems that account for the influence of longitudinal slope on vehicle performance. This study presents a preliminary servo-loop speed control algorithm for connected automated vehicles, enabling smooth, precise, and cost-effective speed tracking (CAVs). The proposed controller focuses on utilizing this readily available future slope information to achieve better speed tracking performance, in contrast to methods that ignore the future road slope and target speed information. It incorporates the future slope and target speed to solve an augmented optimal control problem and obtain the optimal control law in an analytical form. According to experimental results, lower speed tracking errors, more gentle operations, and smoother brake/throttle behaviors are the three main advantages of the proposed control [69]. This study suggests a hierarchical design of optimal car-following control where the system is split into two subsystems with different dynamic properties. It has been evaluated through tests that the proposed optimal control system outperforms a factory-installed adaptive cruise controller in terms of car-following performance [70]. This study suggests a position estimation algorithm that considers road slope and combines GPS and onboard sensor data. This approach makes use of both lateral and longitudinal gradients. The experimental results demonstrate that the road slope-assisted position algorithm outperforms a planar vehicle model-based position estimation algorithm in mountainous terrain in terms of accuracy and reliability [71].

### B. Lateral Slope Control

A few studies are found on road lateral-slope estimation. Most studies on vehicle stability are based on known lateral slope information [46]. Lateral stability control is mostly from the improvement of a model-based integrated controller, which is based on Model Predictive Control (MPC) [72]. Other studies integrate continuous damping control and electronic stability control to prevent rollover by improving lateral stability [73]. Similarly, a non-linear vehicle model is used in situation-based scenarios to reduce rollover by designing a yaw stability control [74]. Some studies calculate the likelihood of a vehicle rollover, also called the lateral load transfer ratio. A predictive-LTR rollover system also improves the stability based on the driver's steering input and sensor signals [75]. In another study, a combination of supervisory, lower-level, and upper-level controllers improves maneuverability and lateral stability for four-wheel-drive electric vehicles [76]. No studies are found on lateral slope control on online estimation. Since most cars are human-driven and not fully autonomous currently, very few studies are done on them. However, the fully autonomous era has already started, and the companies are already building it. Therefore, it is very crucial for an autonomous system to have knowledge about lateral slope and stability.

### V. FUTURE PROSPECT

Most studies assume some other medium provides the slope; however, in the future, the vision-based slope can be included. RSC can be introduced in two different ways into the autonomous modules of delivery vehicles. One is before the low-level controller module of the vehicle and after the motion planning controller. Another one is using RSC within the motion planning controller. Fig. 11 will demonstrate one way of the RSC methods into the controller. The lateral controller of motion planning methods can be the pure pursuit method or Stanley lateral controller, or MPC. The longitudinal controller is simply the velocity controller. On the other hand, the RSC can be a model-free or model-based controller such as PID, fuzzy logic, LQR, MPC and so on. In the second approach, the RSC can be introduced within the motion planning controllers, and the output of the motion planning controller is directly set to the actuator.

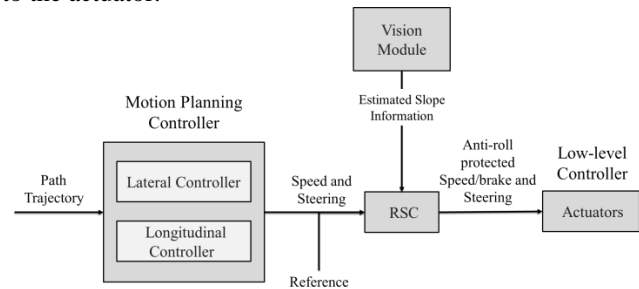

Fig. 11: Roll Stability Control Approach for a delivery vehicle.

### A. PID-based RCS

PID control is a widely used technique for system control in which the system can be brought to the desired set-point by changing just one variable. It compares a given set-point with the state of the system and calculates the difference as an error. The P (proportional), I (integral), and D (derivative) blocks of the PID controller receive this error as input. These blocks have various effects on the error, and the final output is the sum of the blocks. Proportional-Derivative-Integral, or PID, is an abbreviation for the three mathematical procedures used to choose the output to the control system. The standard equation (17) stands as follows:

$$\text{PID output} = K_p e(t) + K_i \int e(t)\mathrm{d}x + K_d \frac{d}{dt} e(t) \qquad (17)$$

Where $e(t)$ is the difference between estimated slope and set slope value, $K_p$, $K_i$ and $K_d$ are the PID gains. The following Fig. 12 describes how these P, I, and D blocks function can be used to control the stability.

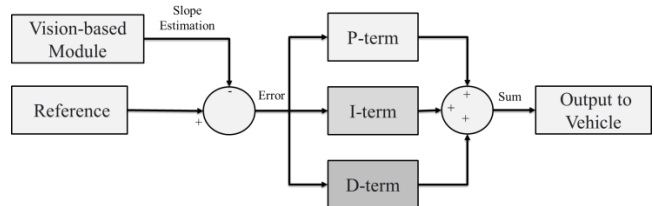

Fig. 12: PID-based RCS

### B. Fuzzy Logic-based RCS

The foundation of fuzzy control systems is a mathematical system that evaluates analogue input values in terms of logical variables and gives output based on the logic. A fuzzy system is a storehouse of fuzzy knowledge that can make decisions on data without using exact Boolean logic. The rule-base has the following structure and represents the expert knowledge as a collection of fuzzy membership functions: If (certain criteria are met), then (results are implied). The fuzzy concept for RCS can be like the following Fig. 13:

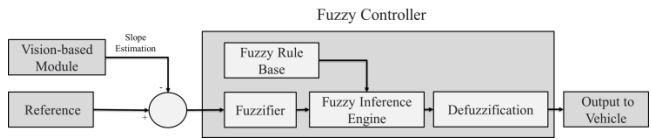

Fig. 13: Fuzzy Logic-based RCS

## C. Sliding Mode-based RCS

In some cases, PD control is recommended over PID control to prevent overshoot. However, PD control results in steady-state inaccuracy since the system is always being perturbed. The study of a sliding mode controller with a saturated integrator addresses both overshoot and steady-state error [76]. Sliding mode control is a non-linear control technique that forces the variable to slide along its normal state by imposing a discontinuous control signal. There is no continuous function of time for the state-feedback control law. Instead, depending on where it is in the state space, it can change from one continuous structure to another. Fig. 14 presents a single SMC controller.

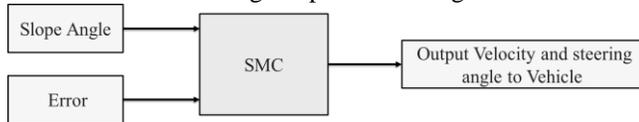

Fig. 14: Sliding mode controller diagram

## D. RSC along with the LQR lateral controller

Typically, the controllers contain a lateral controller for steering commands and a longitudinal controller for throttle and brake commands. The slope error information can be used in the lateral controller together in order to also control the roll stability as well. LQR is a model-based controller that minimizes error by utilizing the vehicle's status. This controller's dynamic model is a simple bicycle model with side slip. The path curvature information is used by the open loop to cancel the constant steady state heading error. However, the closed-loop LQR can be as follows if the slope estimation is considered:

- Lateral Error
- Lateral Error Rate
- Slope Error
- Slope Error Rate

The main idea in LQR control design is to minimize the quadratic cost function of a given dynamics of a system [77]. The cost equation (18) could be as follows:

$$\text{cost} = \int_0^\infty (x^T Q x + u^T R u) dt \tag{18}$$

where $Q$ is an $n \times n$ matrix that expresses the penalty and $R$ is an $n \times n$ matrix that expresses the effort. Solving the LQR problem will return the gain matrix that will produce the lowest cost.

## E. MPC and RSC

MPC makes predictions about the system's future behaviour by using a model of the system. MPC uses an online optimization technique to discover the best control maneuver and match the expected output with the reference [78]. Since it takes the vehicle model into consideration and can optimize different cost functions, it is more accurate and powerful than PID. The main drawback is slower runtime. The cost function should contain the deviation from the reference. The deviation from the reference should be included in the cost function. The primary idea behind MPC is to utilize a system model to anticipate the system's future evolution. Fig. 15 presents a sample diagram of the predictive model controller [79].

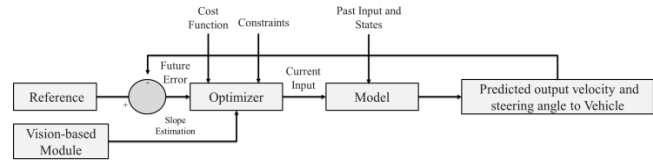

Fig. 15: Diagram of MPC

## VI. Conclusion

In this review, three different problems have been tackled and reviewed. The first one was the road slope estimation methods. Different estimation methods were discussed, and the basic slope estimation process was summarized. Each step within this process was reviewed and analyzed individually. Other methods that do not fall in this category were also discussed. Real-time slope estimation methods were added too for completion. Next, the vehicle model for the control system was discussed. It consisted of reviewing and comparing different tyre models and describing a complete longitudinal vehicle model. In the end, various vehicle control systems that use road slope (predominantly longitudinal slope) to control the vehicle's dynamics were studied. These controllers were mainly used for the purposes of fuel management, speed control, vehicle state estimation and others.

One puzzling point was that almost no research was done on rollover prevention systems that use lateral slopes either through vision-based or real-time systems. This is a big problem as an anti-roll system is a must for vehicles to prevent accidents while cornering or sudden slope changes. This is especially true for autonomous cars and bots. For this reason, in the future prospects section, we have presented a basic idea for anti-roll controllers that make use of slope using various control methodologies. This data can be used for further research and development within this domain.

## Acknowledgment

We acknowledge the support from Dr. Xianke Lin's startup fund.

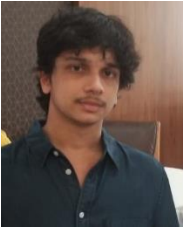

**Gautam Shattey** is currently pursuing his bachelor's degree in Mechanical Engineering from the Indian Institute of Technology Roorkee in India. His current research includes autonomous delivery vehicles, computer vision, machine learning, vehicle dynamics and controller development.

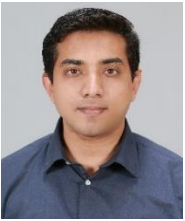

**Sabir Hossain** received B.Sc. in Mechanical Engineering from the Chittagong University of Engineering, Bangladesh, in 2015 and a Master of Engineering from Kunsan National University, South Korea, in 2020. Currently, he is pursuing Ph.D. at Ontario Tech University, Canada, under the supervision of Dr. Xianke Lin. His research includes designing, developing and creating autonomous delivery systems (Mainly Software). His main research interest is localization, perception and planning of Autonomous delivery systems.

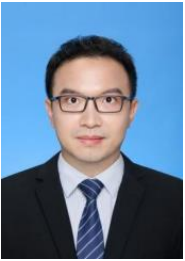

**Chuan Hu** received the B.E. degree in automotive engineering from Tsinghua University, Beijing, China, in 2010, the M.E. degree in vehicle operation engineering from the China Academy of Railway Sciences, Beijing, in 2013, and the Ph.D. degree in mechanical engineering from McMaster University, Hamilton, ON, Canada, in 2017. He was a PostDoctoral Fellow with the Department of Systems Design Engineering, University of Waterloo, Waterloo, ON, Canada, from July 2017 to July 2018, and a Post-Doctoral Fellow with the Department of Mechanical Engineering, University of Texas at Austin, Austin, TX, USA, from August 2018 to June 2020, and an Assistant Professor at the Department of Mechanical Engineering, University of Alaska Fairbanks, Fairbanks, AK, USA, from June 2020 to June 2022. He is currently a tenure-track Associate Professor at the School of Mechanical Engineering, Shanghai Jiao Tong University, Shanghai, China. His research interests include decision-making, path planning, motion control and estimation of autonomous vehicles, vehicle system dynamics and control, humanvehicle interaction, trust dynamics, and robust and adaptive control.

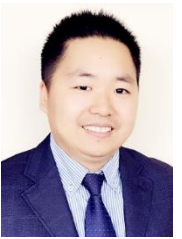

**Xianke Lin** (M'19) received his Ph.D. in Mechanical Engineering from University of Michigan, Ann Arbor, in 2014. He finished his B.S. from Zhejiang University, China, in 2009. He has extensive industrial experience at Fiat Chrysler Automobiles, Mercedes-Benz R&D North America, and General Motor of Canada. Currently, he is an Assistant Professor in the Department of Automotive, Mechanical and Manufacturing Engineering, Ontario Tech University, Oshawa, Canada. His research activities have concentrated on hybrid powertrain design and control strategy optimization, Multiscale/Multiphysics modeling and optimization of energy storage systems.